\documentclass[prb,twocolumn,aps,showpacs]{revtex4-1}

\usepackage[dvips]{graphicx}
\usepackage{amssymb,amsfonts,amsmath}

\begin{document}

\title{(Ir-)Reversibility and thermal equilibrium in magnetic domain pattern formation in ultra-thin ferromagnetic films}

\author{N. Saratz}
\author{U. Ramsperger}
\author{A. Vindigni}
\author{D. Pescia}

\affiliation{Laboratory for Solid State Physics, ETH Zurich, 8093 Zurich, Switzerland}



\begin{abstract}

We investigate the details of pattern formation and transitions between different modulated phases in ultra-thin Fe films on Cu(001). At high temperature, the transitions between the uniform saturated state, the bubble state and the striped state are completely reversible, while at low temperature the bubble phase is avoided. The observed non-equilibrium behavior can be qualitatively explained by considering the intrinsic energy barriers appearing in the system due to the competition between the short-ranged exchange and the long-ranged dipolar interactions. Our experiments suggest that the height of these energy barriers is related to the domain size and is therefore strongly temperature dependent. 
\end{abstract}
\maketitle


\section{Introduction}

Ultra-thin ferromagnetic iron films on the Cu(001)-surface are magnetized perpendicularly to the film plane\cite{Allenspach_PRL_1992,Poppa_JApplPhys_2002,Bauer_JMMM_2004}. Due to the competition between the short-ranged ferromagnetic exchange interaction and the long-ranged dipolar interaction, the magnetization breaks up into magnetic domains which are organized into more or less regular patterns\cite{Yafet_PRB_1988,Kashuba_PRB_1993}. The formation of patterns as a result from competing interactions on different length scales is observed in many physical and chemical systems ranging from type-I-superconductors in the mixed state\cite{Menghini_PRB_2007} or amphiphilic molecules at the air-water interface\cite{Seul_Science_1995} to thermal convection\cite{Hebert_PRE_2010}. Which pattern occurs depends on various parameters, in the case of magnetic films e.~g.\ on the temperature $T$ and the magnetic field $H$ applied perpendicularly to the film plane. In two dimensions, regular patterns include parallel stripes of alternating magnetization and circular bubble domains in a homogeneously magnetized background. Both types of patterns were observed in magnetic garnet films with a thickness of several micrometers\cite{Kooy_PhilResRept_1960,Druyvesteyn_PhilResRept_1971,Cape_JApplPhys_1971,Garel_PRB_1982,Seshadri_PRB_1992}. However, transitions between the stripe and bubble phase could only be induced if a small AC-field was applied to excite the system.
In ultra-thin magnetic films metastable bubble patterns could be observed after applying a strong, almost in-plane, magnetic field pulse\cite{Choi_PRL_2007}. Also in the switching of magnetic multilayer systems metastability and out-of-equilibrium states are the rule rather than the exception\cite{Davies_PRB_2004}.

Recently, \emph{equilibrium} transitions between stripe, bubble, and uniform patterns were observed at high enough temperature in ultra-thin Fe films on the Cu(001)-surface\cite{Saratz_PRL_2010} and the phase diagram in the $T$-$H$-plane was experimentally determined. For convenience, the phase diagram is reproduced schematically in Fig.~1. In the present paper we investigate the details of the transformations between the different patterns by following paths of constant field or constant temperature in the phase diagram. While at high temperatures all processes are completely reversible, at low temperature we observe hysteretic behavior in both types of paths. Moreover, we find that the bubble pattern is systematically avoided at low temperatures. These non-equilibrium aspects result from energy barriers which intrinsically arise from the competition between the short- and long-ranged interactions\cite{Tarjus_JoPCondMat_2005,Schmalian_PRL_2000,Muratov_PRE_2002}.  They occur also under ideal conditions, i.~e.\ in absence of pinning to structural defects. The energy landscape of the system is explored in a ground-state continuum model, which qualitatively explains the observed behavior at all temperatures.

The structure of this paper is as follows. In section II we discuss details of the sample preparation procedure, the experimental set up, and the general properties of our samples. In section III we introduce the important quantities that characterize a given sample. In section IV we image the transformations of the domain patterns along different paths in the $T$-$H$-plane and discuss their (ir-)reversibility. In section V we present details of the transformations in the vicinity of the phase boundaries between different patterns. Finally, in section VI, we summarize the essential conclusions drawn from the experimental observations.
In appendices A and B, we present an extended report on the ground-state energy computations, some results of which are used to interpret experimental observations.

\begin{figure}
\includegraphics{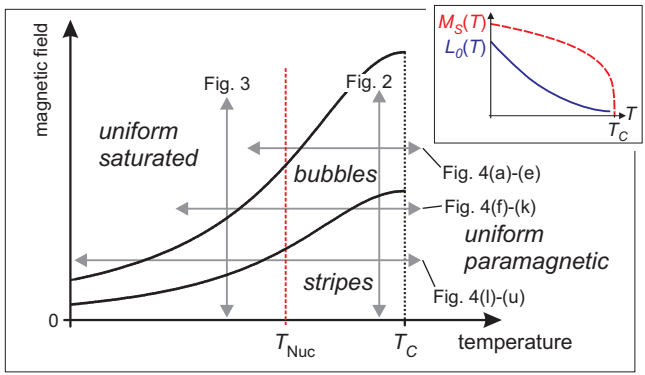}
\caption{(color online) Schematic pattern phase diagram for ultra-thin Fe films with perpendicular magnetization according to Ref.~\onlinecite{Saratz_PRL_2010}.
The gray double arrows illustrate the paths followed for the measurement of figures 2 to 4 as indicated. The vertical dashed lines mark characteristic temperatures $T_\mathrm{Nuc}$ and $T_C$ as discussed later in the text. The inset shows schematically the temperature dependencies of the important parameters $L_0$ (blue-continuous curve) and $M_S$ (red-dashed curve), see the text.} 
\end{figure}

\section{Sample preparation and experimental setup}

Our samples are grown by molecular beam epitaxy  at room temperature in ultra-high vacuum (UHV, base pressure below $1.5\cdot10^{-10}$~mbar) with a typical rate of 0.1 to 0.2 atomic layers per minute.
The substrate is a copper single crystal with the (001)-surface oriented to better than $0.1^\circ$ and polished to $<0.03~\mu$m roughness depth\cite{MaTecK}.
The substrate is cleaned by several cycles of Ar-ion bombardment (argon pressure $2.5\cdot 10^{-5}$~mbar, ion energy 1~keV) and annealing at 700~K for 50 minutes.
The chemical cleanliness and film thickness are determined using Auger electron spectroscopy, the crystallinity of the sample is verified by low-energy electron diffraction. The film thickness $d$ is measured with an accuracy of $\pm0.02$ nominal atomic monolayers (ML). After preparation, the sample is transferred to the measurement chamber (base pressure below $3\cdot10^{-11}$~mbar) in the same UHV system.
We measure the local magnetization of the sample with a lateral resolution of 50~nm using Scanning Electron Microscopy with Polarization Analysis of the secondary electrons (SEMPA)\cite{Koike_APL_1984,Scheinfein_RevSciInstr_1990,Allenspach_IBMJResDevelop_2000}.
By scanning the beam of a scanning electron microscope over the sample and analyzing the spin polarization of the secondary electrons generated at the electron beam spot, we obtain a spatial map of the local magnetization vector.
A magnetic field of up to 0.4~mT can be applied to the sample while measuring and the sample temperature can be varied from 20 to 400~K.
The magnetic field can be controlled on the level of 1~$\mu$T and the temperature is stabilized to less than 0.1~K.
For electron-optical reasons, in our experiment the field is applied at an angle of 45${}^\circ$ with respect to the film normal.
However, within our experimental resolution we do not observe any in-plane component of the magnetization nor do we see an influence of the in-plane component of the field on the domain pattern when we rotate the sample about its vertical axis.
We conclude therefore that for the very weak fields applied in the present experiments, only the out-of-plane component of the magnetic field affects the sample. 
Because the Fe films are very susceptible to weak magnetic fields along their normal, it is important to eliminate the perpendicular component of any unwanted field (e.~g.\ stray fields from the magnetic lenses of the electron microscope or the earth magnetic field).
This is achieved by tuning the applied field appropriately and zero-field is defined as the situation with no imbalance between up- and down magnetized domains.
The magnetic field values indicated throughout this paper refer to the effective perpendicular component of the field at the sample. In the images, the out-of-plane component of the magnetization is indicated by a gray scale with black and white corresponding to opposite signs of the magnetization.

\section{Characteristic quantities}
In zero field, we observe a characteristic temperature $T_C$ at which the contrast resulting from the magnetic domains disappears in the SEMPA images\cite{Portmann_PRL_2006}.
This sample-specific $T_C$ sets the temperature scale for each individual film. 
By introducing a reduced temperature $T/T_C$, the temperature dependence of the domain patterns can be compared between different samples with slightly different film thickness (ranging from 1.9 to 2.3~ML) and different $T_C$ (ranging from 320 to above 360~K). As a function of the film thickness, $T_C$ reaches a maximum for 2.15~ML, decreasing rapidly for thinner films and more slowly for thicker films\cite{Portmann_Nature_2003}.
In zero field, the domain pattern consists of parallel stripes of width $L_0$ with alternating magnetization $\pm M_S$ as expected\cite{Allenspach_PRL_1992}. Throughout the temperature range considered in this work, we observe a strong decrease of $L_0$ with increasing temperature\cite{Portmann_PRL_2006,Vindigni_PRB_2008}, see the inset in Fig.~1. As discussed in appendix A, this stripe width in zero field, $L_0$, determines the domain size at all values of the applied field and is therefore an important quantity characterizing the sample at any given temperature.
If a magnetic field is applied, the domains with the appropriate sign of the magnetization are energetically favored and their area grows at the expense of the domains carrying the opposite sign. The pattern acquires an asymmetry $A=(f_\uparrow - f_\downarrow)/(f_\uparrow + f_\downarrow)$, where $f_{\uparrow(\downarrow)}$ is the area occupied by up (down) magnetized domains. The zero-field pattern with alternating stripes of equal width has $A=0$ and the up or down magnetized saturated states in the limit of high fields have $A=\pm 1$. For intermediate values of $A$ at intermediate magnetic fields one expects that the minority, down-magnetized domains form a hexagonal lattice of (almost) circular (bubble) domains in a homogeneous up-magnetized background rather than a striped pattern \cite{Ng_PRB_1995}.
Because the weak magnetic fields considered in this study only change the domain configuration but not the local magnetization inside the domains, the latter is equivalent to the saturation magnetization, $M_S$. The total magnetization of the sample for any value of $A$ is then given by the product $A\!\cdot\!M_S$. Over most of the temperature range $M_S$ is weakly temperature dependent, except close to $T_C$ where it drops sharply, see the inset in Fig.~1.
The magnetic field scale for a given sample at a given temperature is determined by its critical magnetic field $H_C$, defined as the field above which the uniform state $|A|=1$ has the lowest energy. It turns out that the asymmetry $A$ is proportional to the reduced field $H/H_C$ over most of the magnetic field range\cite{Saratz_PRL_2010}. Moreover, $H_C$ can be expressed through the saturation magnetization $M_S$ and the stripe width \emph{in zero field} $L_0$, $H_C\propto M_S/L_0$, as discussed in appendix A, Eq.~\eqref{eq_hml}. Together with the experimental curves for $M_S(T)$ and $L_0(T)$\cite{Vindigni_PRB_2008, Portmann_PRL_2006, Saratz_PRL_2010}, this relation allows us to trace the phase boundary separating the uniform and the bubble states in Fig.~1,
\begin{equation}
\label{eq-hcmain}
 H_C(T)= \frac{4}{e^2}M_S(T)\frac{d}{L_0(T)}\,.
\end{equation}
Inserting typical values for $d=2.0\,\mathrm{ML}=0.36\,\mathrm{nm}$, $\mu_0 M_S=2\,\mathrm{T}$ and $L_0 = 1\,\mu\mathrm{m}$ corresponding to our experimental system a few K below $T_C$, we obtain $\mu_0 H_C = 390\,\mu\mathrm{T}$, in good agreement with the typical saturation fields observed in the experiments, see the next sections and Ref.~\onlinecite{Saratz_PRL_2010}.

\section{Domain patterns vs. field and temperature}

\subsection{Constant temperature}

\begin{figure}
\includegraphics{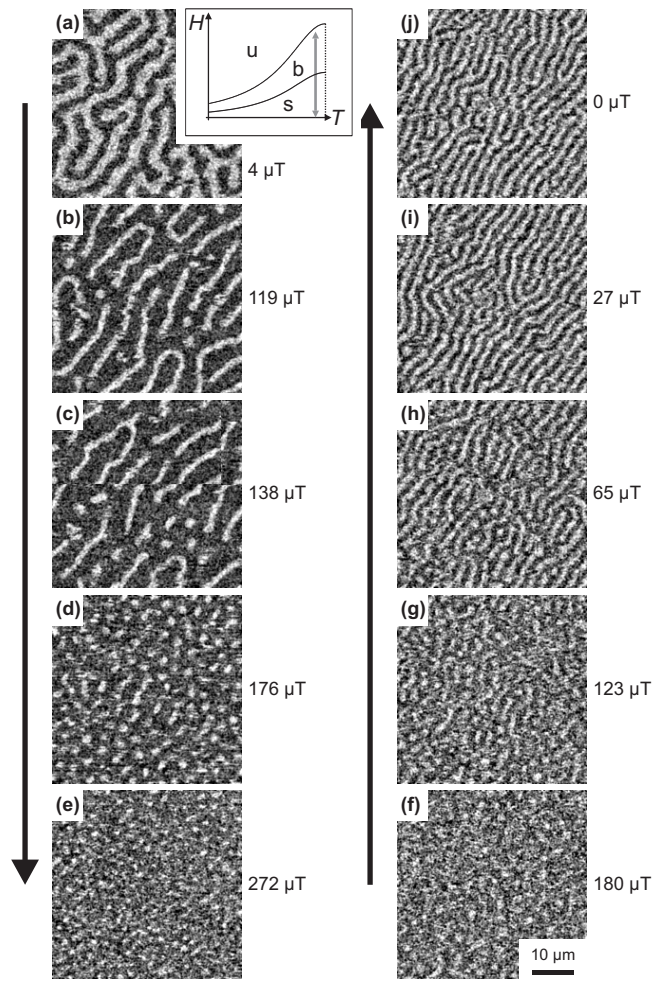}
\caption{Bubble formation upon changing the magnetic field at constant temperature. The absolute value of the magnetic field for each image is indicated.
Images \textbf{(a)--(e)}: increasing magnetic field at $T/T_C$=0.994 ($T$=350~K, $d$=2.00~ML), \textbf{(f)--(j)}: decreasing magnetic field at $T/T_C$=0.996 ($T$=339~K, $d$=1.93~ML).
All image sizes are 45~$\mu$m by 45~$\mu$m, the inset illustrates the path followed in the phase diagram.} 
\end{figure}

\begin{figure}
\includegraphics{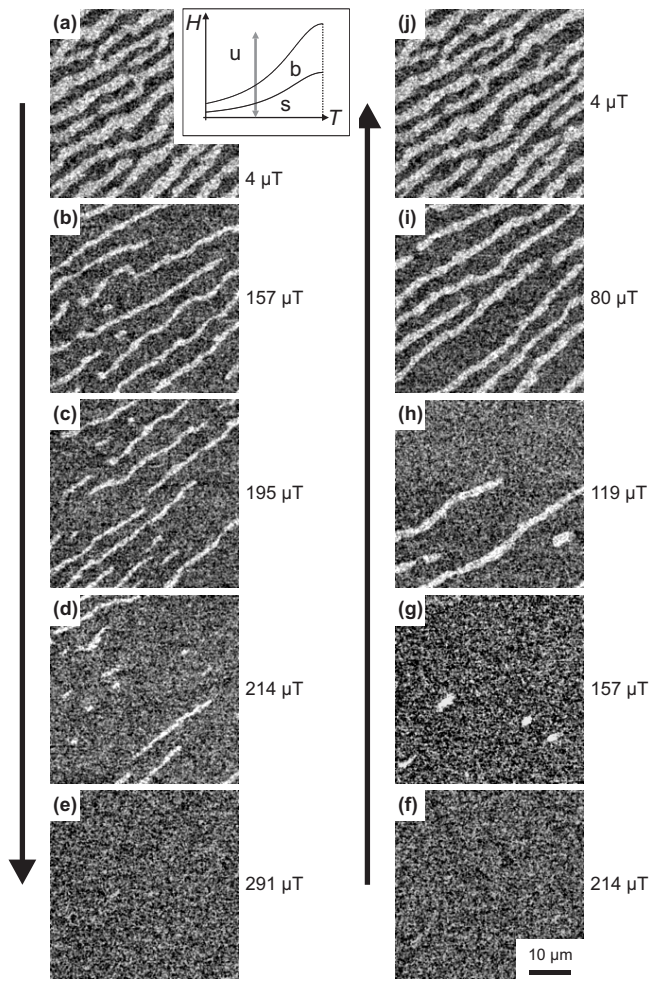}
\caption{Domain patterns upon changing the magnetic field at constant $T$=317~K ($T/T_C$=0.970).
The absolute value of the magnetic field is indicated for each image. Images \textbf{(a)--(e)}: increasing field, \textbf{(f)--(j)}: decreasing field.
Note that images (f)--(j) have been recorded before images (a)--(e) with (j) and (a) being the same image. All image sizes are 45~$\mu$m by 45~$\mu$m, the inset illustrates the path followed in the phase diagram.} 
\end{figure}

The most obvious way to observe a transition from stripe to bubble domains is to apply a magnetic field at constant temperature.
For $T$ close to $T_C$, this process is shown in Fig.~2. We start from a striped pattern at almost zero field, image~2(a), with white and black regions almost balanced. In image~2(b), the external field results mainly in a compression of the (white) minority stripes.
In this step splitting of selected domains has taken place and a few domains have acquired a bubble-like shape.
This trend is continued in image~2(c), where the numbers of stripe-like and bubble-like domains are roughly equal.
In image~2(d) virtually all stripe domains have broken up and the domain pattern consists of a disordered array of white bubbles immersed in a black background.
Upon increasing the magnetic field further, the bubble diameter decreases but the bubble density remains approximately constant.
The transition to saturation is not observed in this case because imaging is not feasible with our experimental setup in the fields required to saturate the sample at this temperature.
Images 2(f) to 2(j) show the domain transformation in decreasing magnetic field, still at high temperature.
The barely visible bubble domains of image~2(f) partially merge, leading to randomly distributed, slightly elongated domains in image~2(g).
These merge further and in image~2(h) only a few bubble domains are left and the pattern clearly has acquired a stripe character.
Finally, in images~2(i) and 2(j) the stripe pattern is restored completely.

Note that the periodicity of the domain pattern, i.~e.\ the typical distance between neighboring bubbles or stripes with the same sign of the magnetization, is essentially unaffected by the magnetic field.
This observation is in agreement with ground state energy calculations, see Fig.~8 and the appendix A.
Since the periodicity of the pattern does not change, the transition from black-and-white stripes to white bubbles on a homogeneous black background implies that the total number of white domains increases and \emph{splitting} of the white domains must occur.
Alternatively, we could also state that the black stripes have to \emph{merge} to form the homogeneous background.
Merging and splitting are therefore completely equivalent -- which process occurs depends only on whether one describes the black or the white domains.
For the sake of simplicity from here on we concentrate on describing the white domains which constitute the minority domains for the direction of the magnetic field applied here.
As discussed in more detail in the appendix B, splitting of domains implies the crossing of an energy barrier because the energy cost associated with temporarily increasing the length of the domain walls is not completely compensated by the gain in dipolar energy. The existence of this energy barrier is an intrinsic property of the pattern forming system and has nothing to do with the extrinsic energy barriers responsible for the pinning of the domain walls at structural defects such as atomic steps on the substrate or the sample border.
The fact that the sequences 2(a)--(e) and 2(f)--(j) are reversible indicates that for comparatively high temperatures merging and splitting readily occur, and the system reaches its equilibrium stripe, bubble or uniform state in agreement with the phase diagram of Ref.~\onlinecite{Saratz_PRL_2010} (Fig.~1). Moreover, pinning at minor structural defects seems to play no role in this case. 



At lower temperatures (Fig.~3) domain splitting is more difficult because the system has less thermal energy to overcome the energy barrier and additionally, the energy barrier itself increases as discussed in the appendix B. There we suggest that the 
height of the energy barrier which impedes domain splitting or merging is proportional to the characteristic domain size $L_0$. This implies that the energy barrier also has the same temperature dependence as $L_0(T)$, meaning that the barrier strongly increases for decreasing $T$. Temperature-dependent energy barriers have been suggested in a frustration-based approach to supercooled liquids in the context of glass-forming systems\cite{Tarjus_JoPCondMat_2005}.
The fact that splitting is forbidden implies that the total number of domains in the sample can only increase by nucleation of new domains. Therefore the response of the system to the applied field is fundamentally different from the equilibrium situation of Fig.~2.
Applying a weak field to the stripe pattern of image 3(a) results in a lateral compression of the stripe domains, see image~3(b), in agreement with the expectations from the ground state calculations and the equilibrium situation.
For larger fields, the system can not reach the equilibrium bubble state because this would require domain splitting and in consequence the system is locked in a metastable stripe pattern, image~3(c).
Nevertheless, the stripes successively disappear, image~3(d), finally giving way to the uniform saturated state, image~3(e), which is unavoidable at large enough fields.
Because a large energy barrier prohibits the lateral collapse of the stripes, see the appendix B, a stripe can only disappear by contracting along its length.
This is not possible as long as the ends of the stripe are pinned at some defect or the sample border. At a high enough field, however, we expect that one end of the stripe will unpin, followed by a rapid contraction of the stripe until it is reduced to a single circular bubble domain. Such residual bubble domains can be observed in images~3(b), 3(c), and 3(d).
As it is pointed out in the appendix, the energy barrier that stabilizes the isolated bubble domain disappears above a critical field $H_\mathrm{collapse} = 1.35 H_{C}$, where $H_C$ is the equilibrium saturation field introduced in section III. 
Therefore, for $H>H_\mathrm{collapse}$ the bubble inevitably collapses, without crossing an intrinsic energy barrier. In consequence, this process is allowed also at low temperature, at least as long as domain wall motion in general is possible.
According to this discussion, the bubble phase is systematically avoided at low temperature, even in the ideal system, in absence of pinning at structural defects. The fact that the images in Fig.~3 have been recorded only 9~K below $T_C$, illustrates the difficulty of observing the equilibrium bubble pattern at constant temperature in this system.



When reducing the magnetic field from the saturated state at low temperature, image~3(f), only a small number of domains nucleate, image~3(g), probably at some structural defects that are not resolved in the image\cite{Saratz_JPhysD_2007}.
Nucleation of a domain -- in contrast to its collapse -- is always hindered by an energy barrier, see the appendix B, although a local modification of the film properties at a structural defect may lower this barrier and promote domain nucleation\cite{Cape_JApplPhys_1971}.
Note that image~3(g) has been recorded at the same magnetic field as image~3(b).
The initial seed domains in image~3(g) expand along one direction, image~3(h), and upon lowering the field further, more stripe domains enter the field of view, image~3(i), until the weight of up and down magnetized domains is essentially equilibrated in image~3(j).

In summary, while at high temperature, in close proximity of $T_C$, the observed domain pattern closely follows the equilibrium phase diagram of Ref.~\onlinecite{Saratz_PRL_2010},
already a few Kelvins below $T_C$ the domain pattern is dominated by metastability. The transition from the striped to the saturated state in increasing field does not proceed via the bubble phase but by direct collapse of the stripe domains\cite{Davies_PRB_2004}.


\begin{figure*}
\includegraphics{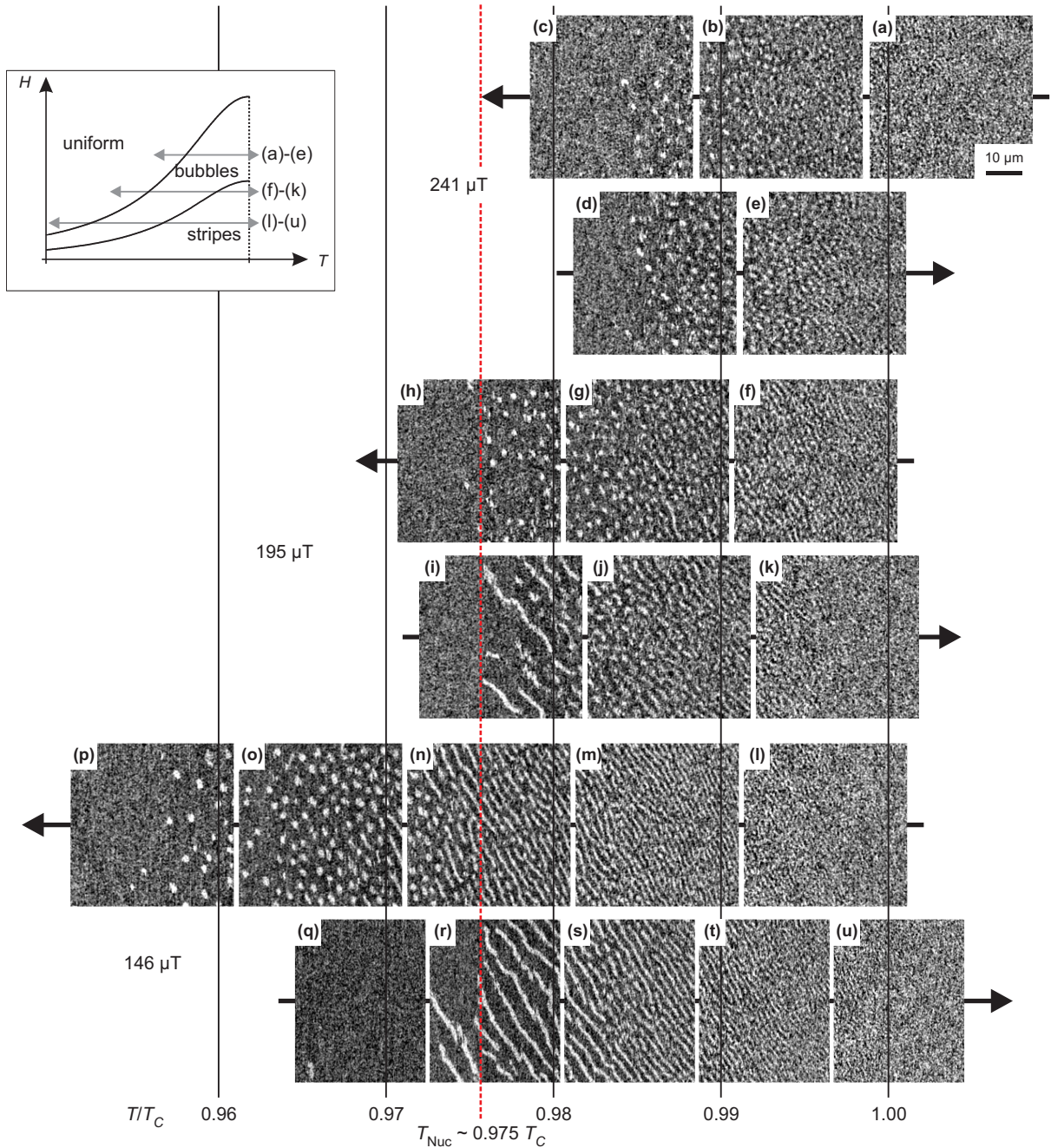}
\caption{(color online). Bubble formation upon changing the temperature in constant field.
The images (a)--(p) were acquired at a constant cooling/heating rate of 0.5 K/minute, leading to a continuous temperature variation of 3.2~K within each image.
Image \textbf{(a)--(c)}: cooling in 241~$\mu$T, \textbf{(d), (e)}: heating in 241~$\mu$T, \textbf{(f)--(h)}: cooling in 195~$\mu$T, \textbf{(i)--(k)}: heating in 195~$\mu$T, \textbf{(l)--(p)}: cooling in 146~$\mu$T, \textbf{(q)--(u)}: heating in 146~$\mu$T at a rate of 0.4~K/min. The images (q)--(u) have been scaled by a factor of 0.8 along the horizontal direction to fit on the same temperature scale.
All image sizes are 45~$\mu$m by 45~$\mu$m.
The film thickness is $d$=2.22~ML, all images were measured on the same sample. The vertical lines indicate the reduced temperature scale and the inset illustrates the paths followed in the phase diagram.}  
\end{figure*}

\subsection{Constant applied field}

As is evident from the phase diagram of Fig.~1, transitions between the different phases can also be induced by varying the temperature while keeping the applied field constant. In Fig.~4 we present the domain patterns observed by following such paths of constant field, as indicated in the inset.
The images of Fig.~4 are obtained by scanning the electron beam over the sample while the temperature is varied at a fixed rate.
With a cooling/heating rate of $\mp$0.5~K/min and acquisition of each image taking 400~seconds, each one of the images~4(a)~to~4(p) comprises a temperature variation of 3.2~K.
A different position in the image is therefore equivalent to a different sample temperature.
The temperature scale is indicated in terms of the reduced temperature by the vertical lines in Fig.~4 and the images are placed accordingly.
Note that images~4(q) to 4(u) have been acquired at a slightly lower heating rate of 0.4~K/min and have been scaled by a factor of 0.8 along the horizontal axis in order to fit on the same temperature scale.

For high magnetic fields [241~$\mu$T, images~4(a)--4(e)], no stripes are observed, in agreement with the phase diagram of Fig.~1. When cooling from above $T_C$, Fig.~4(a), bubble domains appear in image~4(b) whose density decreases monotonously with decreasing temperature until a homogeneous, saturated state is reached, Fig.~4(c).
Since for this field the transitions occur at high temperature, the process is completely reversible, as can be seen in images~4(d) and 4(e) recorded upon heating. We can therefore be sure that the system is at equilibrium, as in Fig.~2. 

For intermediate values of the magnetic field, images~4(f)--4(h), the sequence of phases upon cooling is uniform paramagnetic $\to$ stripes $\to$ bubbles $\to$ uniform saturated.
The transition from stripes to bubbles occurs at a high temperature of about 0.99~$T_C$, see images~4(f) and~4(g), at which domain splitting occurs and the process is equivalent to the observations of Fig.~2.
Upon decreasing the temperature further, the density of the domains decreases by successive collapse of the bubbles, until the saturated state is reached in image~4(h), at a clearly lower temperature than for the high field in image~4(c).
Remember that collapse of the bubble domains is allowed also at low temperatures.
Upon heating, a different behavior is observed.
The sample remains saturated until a sample-specific temperature $T_\mathrm{Nuc}$ is reached and suddenly, from one scan-line to the next one, stripe domains appear on the sample, image~4(i). As pointed out before, nucleation always implies the crossing of an energy barrier, for which a certain thermal activation is required. In the appendix B we compute the height of this nucleation barrier $E_B$ for a single bubble domain in the ground state. Analogously to the discussion of $H_C(T)$, see Eq.~\eqref{eq-hcmain} in section III, we may translate the ground-state result of Eq.~\eqref{eq-ebappscaling} to finite temperatures by using the experimental results for $M_S(T)$ and $L_0(T)$. For $H=H_C$ corresponding to the equilibrium transition from the uniform to the bubble state we obtain
\begin{equation}
\label{eq-barriermain}
E_B(T) = 0.1904 \mu_0 M_S(T)^2 d^2 L_0(T) \, .
\end{equation}
Eq.~\eqref{eq-barriermain} suggests that the barrier height $E_B$ is proportional to $L_0$. Using the same values as before, $d=0.36\,\mathrm{nm}$, $\mu_0 M_S=2\,\mathrm{T}$ and $L_0 = 1\,\mu\mathrm{m}$, in \eqref{eq-barriermain}, we obtain $E_B/k_B=5680\,\mathrm{K}$, where $k_B$ is the Boltzmann constant. At an absolute temperature of $330\,\mathrm{K}$, we obtain then a probability $\exp\left(-E_B/k_B T\right)=3.3\cdot10^{-8}$, which gives a reasonable nucleation rate if one assumes an attempt frequency of $\mathcal{O}\left(10^{9}-10^{12}\right)\,\mathrm{Hz}$\cite{Novak_JMMM_2005,Ehlers_PRB_2006}. Note that according to \eqref{eq-barriermain}, the main temperature dependence of the nucleation probability is not due to the varying absolute temperature $T$, but rather due to the strong temperature dependence of $E_B(T)$ via $L_0(T)$.  The nucleation temperature $T_\mathrm{Nuc}$ is therefore determined by the \emph{reduced} temperature $T/T_C$ at which $L_0(T)$
(and with it $E_B$) drops below some threshold value. The absolute temperature plays only a minor role.

The stripes in image~4(i) decay to the equilibrium bubble state, Fig.~4(j), as soon as the temperature is high enough to allow for the splitting of domains. At higher temperature the equilibrium transition from bubbles to stripes occurs, and the stripes in turn disappear as $T_C$ is reached, Fig.~4(k).
The transition from saturation to bubbles via transient stripe-like domains is discussed in more detail in the next section in the context of figure~5.

At lower fields, the equilibrium phase sequence is again observed upon cooling.
Because the temperature interval for each phase is wider and the domain size at lower temperature is larger, the phases can be identified more clearly in the images~4(l)--4(p). 
Upon heating, the crossing of the nucleation temperature $T_\mathrm{Nuc}$ can be identified clearly, although a few domains are visible already at lower temperature in images~4(q) and 4(r). In this case $T_\mathrm{Nuc}$ lies above the bubble-stripe transition temperature, compare images~4(n) and 4(r).
Therefore no bubble phase is observed upon heating in low fields, the nucleated stripe pattern is stable and persists upon further heating up to $T_C$.

\section{Details of the pattern transformations}

\begin{figure}
\includegraphics{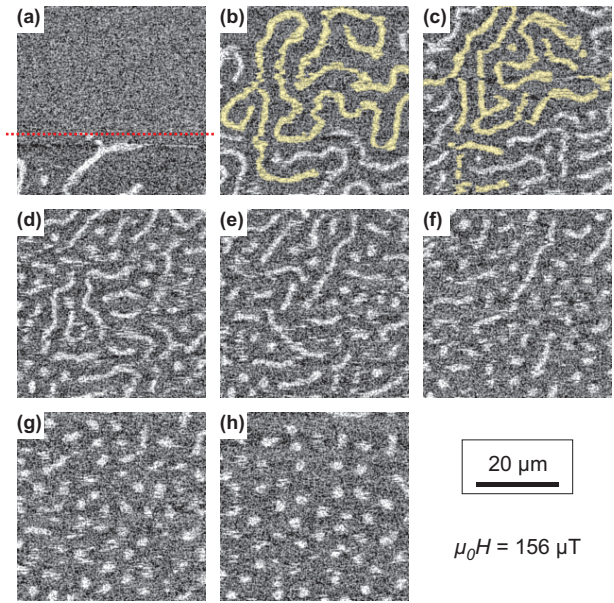}
\caption{(color online). Transition from the saturated state to the bubble state via a transient, string-like domain pattern in a constant magnetic field of 156~$\mu$T. During the acquisition of image (a) the sample is heated from 332.5~K to 335.5~K. In image (b) the temperature settles to 336.5~K and is then held constant for the remaining images. All image sizes are 45~$\mu$m by 45~$\mu$m, the film thickness is $d$=2.00~ML. The slow scan direction is from top to bottom and the acquisition time is 3.5 minutes for each image.}
\end{figure}

At high applied fields and rising temperature the transition from saturation to the bubble pattern proceeds by direct nucleation of circular domains.
At low fields the bubble phase is avoided completely and nucleation directly leads to a striped pattern.
The transition from the saturated state to the bubble state via nucleation of a transient stripe pattern as observed in Fig.~4(i)--4(j) is presented in more detail in Fig.~5.
An initially saturated sample is heated in a constant magnetic field of 156~$\mu$T with a constant rate of +1~K per minute.
When the nucleation temperature marked by the red dotted line in image~5(a) is crossed, suddenly reversed white domains appear on the sample. This process can be understood as follows. The uniformly magnetized state becomes metastable when the equilibrium transition line from bubbles to saturation is crossed upon heating. However, nucleation is prohibited until the associated energy barrier has sufficiently decreased, as discussed in the previous section and the appendix B, and at some temperature $T_\mathrm{Nuc}$, isolated bubble domains nucleate. Since the expansion of an existing isolated domain is not hindered by an energy barrier, these domains grow rapidly (instantaneously, on the time scale of our measurement) until the equilibrium asymmetry $A$ corresponding to the applied field and temperature is reached. This process is in agreement with the observations by Cape and Lehman \cite{Cape_JApplPhys_1971} in thick garnet films.
At this point the temperature ramp is halted and $T$ is kept constant during the remaining images. As can be seen from the contiguous domain marked yellow in image~5(b), the initial domains extend over large areas.
However, in contrast to thick garnet films, in our case thermal fluctuations are strong enough to break up the string-like domains and increase the total number of domains.
This process can be observed already in image~5(c) where the many colored domain segments originate from the single domain marked in image~5(b) and is more pronounced in images~5(d), 5(e), and 5(f).
Finally, in images~5(g) and~5(h) only bubble domains of approximately circular shape are left.

\begin{figure}
\includegraphics{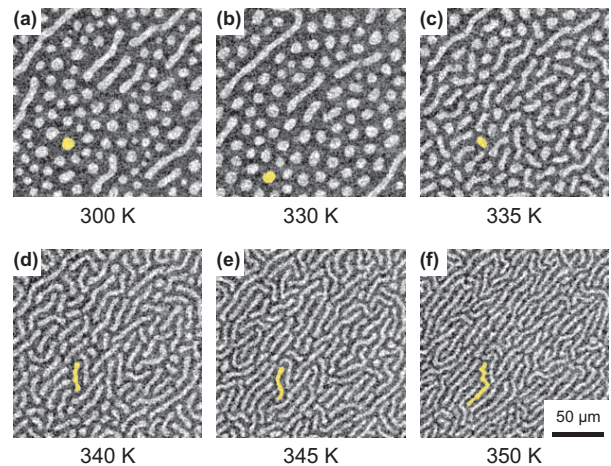}
\caption{(color online). The sequence of SEMPA images shows the domain pattern transformation on heating a bubble domain pattern from 300~K up to 350~K. The magnetic field is 22~$\mu$T. Image size is 182~$\mu$m by 182~$\mu$m for all images and the film thickness is $d$=2.16~ML.}
\end{figure}

The transformation from bubbles to stripes in a constant field of 22~$\mu$T is investigated in Fig.~6. To help following the pattern transformation, one representative domain is highlighted in image~6(a) and the same domain is marked yellow in all images.
Images~6(a) and 6(b) show that at low temperature the domain pattern is mostly unaffected by the temperature change of 30~K.
Both images show a pattern of essentially randomly arranged round domains.
In image~6(b) the size distribution is narrower and the domains have a more circular shape, indicating that the pattern is closer to equilibrium.
From image~6(b) to 6(c) the bubble domains elongate and no preferential direction can be observed for this elongation.
In image~6(d) the aspect ratio of the domains increases further and in image~6(e) the domains start to preferentially align along the same direction.
Note that in spite of the high absolute temperature, no domain merging or splitting occurs due to the low reduced temperature ($T_C>360~K$) and associated large domain size. Therefore the total number of domains remains constant, the transformation from the bubbles to the stripes is achieved only by the growth in length and shrinking in width of the domains. 
This holds true also for image~6(f), in which the order of the stripe pattern increases further.
The transformation from bubbles to stripes in increasing temperature at constant field is driven by two effects. Because the equilibrium periodicity of the pattern $L_0$ decreases, the width of the white domains tends to decrease. However, because $H_C(T)$ increases and $A\propto H/H_C$ , the equilibrium asymmetry $A$ decreases and therefore the area occupied by the white domains tends to increase. The decrease in width and simultaneous increase in area favors an elongated shape of the domains and thus a transformation from bubbles to stripes with increasing temperature, without the need for domains to merge.


\begin{figure}
\includegraphics{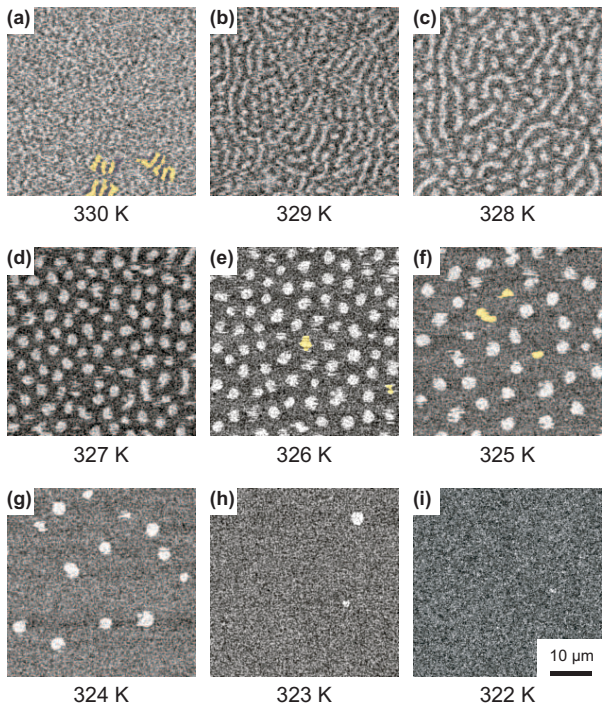}
\caption{(color online). The sequence of SEMPA images shows the domain pattern transformation on cooling from above the Curie temperature $T_C$ to saturation. The magnetic field is 95~$\mu$T and all image sizes are 45~$\mu$m by 45~$\mu$m. The film thickness is $d$=1.91~ML.}
\end{figure}

A detailed measurement of the domain pattern evolution upon cooling from above $T_C$ in a constant applied field is shown in Fig.~7.
The sample is cooled in discrete steps of -1~K and after the temperature is allowed to settle during 3~minutes, the acquisition of the next image is started.
All images show the same field of view. The image~7(a) taken at 330~K $\approx T_C$ is essentially contrastless and shows only small regions of stripe domains marked in yellow and blue.
In image~7(b) a mostly striped pattern can be observed, intermixed with some bubble domains.
The next \mbox{image, 7(c),} shows a mixture of elongated domains and approximately circular bubbles with a larger average domain size than in image7(b), in agreement with the larger $L_0$ at lower $T$.
At a temperature of 327~K, in image~7(d), except for a few all domains have a circular shape and in image~7(e) only circular domains are left.
As it is obvious from the image, the bubbles are not arranged on a hexagonal lattice as one might expect but they are rather disordered as was observed already in Fig.~5(d) and 5(e). 
Since the fast scanning direction of the microscope is from left to right, the fact that domains are moving during scanning manifests as horizontal streaks in the image.
Two domains showing such streaks are marked yellow in image~7(e).
This mobility is also observed in image~7(f), for example in the yellow colored domains.
From images~7(e) to 7(f) and 7(g) the bubble density decreases with decreasing temperature.
In image~7(h) only two reversed domains are left and finally, in image~7(i), the entire field of view is homogeneously magnetized.
As long as domain merger, splitting, and nucleation are not possible, the total number of domains in a sample can only decrease with time, since only domain deformation or collapse are allowed. For decreasing temperature, the increase in the reduced field $H/H_C(T)$ tends to decrease the area occupied by the white domains. This decrease of the area proceeds by contraction of the stripes along their length. Simultaneously, the increase in $L_0(T)$ favors an increase of the domain width and domain separation. Therefore, the temperature dependencies of both, $L_0(T)$ and $H/H_C(T)$ favor a transition from elongated to circular domains with decreasing $T$, analogously to the case of increasing temperature discussed in Fig.~6.


\section{Conclusions}

Our experiments show that in two-dimensional, ultra-thin ferromagnetic films with perpendicular anisotropy thermal fluctuations alone are sufficient to promote the transitions between stripe, bubble, and saturated phases in a narrow temperature interval below $T_C$. However, in general the actual pattern realized at a given magnetic field and temperature depends not only on the equilibrium pattern as deduced from the phase diagram in Fig.~1, but also on the history of the sample and the specific path followed in the $T$-$H$-plane. At constant temperature, the characteristic size of the domain pattern is constant and a transition between the stripe and the bubble pattern necessarily involves merger or splitting of the domains. These processes, as well as nucleation, are inhibited by energy barriers that are inherent to the ideal system and are a direct result of the competition between the short- and long-ranged interactions. Translating the results of our ground-state computations to finite $T$ suggests that the heights of these energy barriers are proportional to the domain size $L_0$ and do therefore strongly depend on temperature, in the same way as does $L_0(T)$. We conjecture that for decreasing temperatures domain merger, splitting and nucleation are suppressed not only because there is less thermal energy available, but mostly because the associated energy barriers increase strongly.
In contrast, on a path of constant applied field $H$ and, for instance, decreasing temperature, both the increase of $L_0(T)$ and the increase of $H/H_C(T)$ promote a transition from a stripe to a bubble pattern. Moreover, in this case the transformation is achieved by changing the shape of the existing domains only, no domain splitting is required and no energy barrier needs to be crossed. As a summarizing rule we may state that below a certain threshold temperature the number of domains in the sample can only stay constant or decrease with time. While a pure rearrangement of the domains may accommodate a transition from stripes to bubbles and vice versa for changing temperature, this rule forbids the transition from stripes to bubbles at constant low temperature.
The pinning to structural defect in the sample and on the substrate plays no important role in the samples and in the temperature range considered in this work. It is therefore tempting to speculate that metastability and the associated disorder in the domain pattern are indeed self-generated by the long-range interaction\cite{Tarjus_JoPCondMat_2005,Schmalian_PRL_2000,Muratov_PRE_2002} and entropy. 


\section*{ACKNOWLEDGEMENTS}

We thank Thomas B\"ahler for technical assistance and ETH Zurich as well as the Swiss National Science Foundation for financial support. 

\appendix

\section{Ground-state phase diagram}

We compute the total energy of the domain patterns in different situations by minimizing the total energy of a slab with thickness $d$ along the $z$-direction and lateral extension $\Lambda\gg d$ in the $x$-$y$-plane. 
In a continuum approach, we may represent the configuration of magnetic domains by a two-valued scalar field $m(\vec \rho)=\pm1$ if we assume that the magnetization points only along $\pm z$ and is homogeneous along the small thickness $d$. With this we have assumed that the domain walls are sharp and the local magnetization at a position $\vec \rho$ in the $x$-$y$-plane is given by $M_S m(\vec \rho)\vec e_z$. The Hamiltonian for this model reads
\begin{eqnarray}
\label{eq-ham}
2J d l_{wall}  \; +\;  \lambda \iint m(\vec \rho) V_d(|\vec \rho-\vec\rho'|) m(\vec \rho')\mathrm{d}^2\rho \;\mathrm{d}^2\rho' \nonumber \\
-\;  h d \int m(\vec \rho)\mathrm{d}^2\rho\, .\qquad
\end{eqnarray}
The first term, corresponding to the exchange interaction with coupling strength $J$ per atom, increases the energy stored in the domain walls by an amount proportional to the film thickness $d$ and the total length of the domain walls in the sample $l_{wall}$. All lengths are measured in units of the lattice constant. The second term represents the dipolar energy (coupling constant $\lambda\doteq \frac{\mu_0}{4\pi} M_S^2$) as computed from the interaction between the magnetic surface charges at the top and bottom surfaces of the slab following Refs.~\onlinecite{Kooy_PhilResRept_1960,Druyvesteyn_PhilResRept_1971,Garel_PRB_1982},
with
\begin{equation}
V_d(|\vec \rho-\vec\rho'|) = \frac{1}{|\vec \rho-\vec\rho'|} - \frac{1}{\sqrt{|\vec \rho-\vec\rho'|^2 + d^2}}\, ,
\end{equation}
as given in more detail in Ref.~\onlinecite{Saratz_PhD_2009}. 
The last term in \eqref{eq-ham} describes the interaction with the external field, $h=\mu_0 H M_S$, and is equivalent to $-d \Lambda^2 \mu_0 H M_S A(H)$, with $A = \frac{1}{\Lambda^2}\int m(\vec \rho)\mathrm{d}^2\rho $ being the asymmetry as introduced in section III.
We consider the following domain configurations:
i) parallel stripes along the $y$-axis, with alternating magnetization $m=\pm M_S$ and alternating width $L\pm\delta$, such that the periodicity of pattern is $L_S=2L$ and the width of the narrower stripes is $w=L-\delta$.
ii) a hexagonal array of reversed ($m=-M_S$) bubbles of radius $R$ and center spacing $L_B$ on a homogeneous background ($m=+M_S$).
iii) the saturated state $m=+M_S$.
For a given set of the parameters $J$, $d$, $\lambda$, and $h>0$ we minimize \eqref{eq-ham} with respect to $L$ and $w$ or $L_B$ and $R$, respectively, and compare the energies of the optimized stripe, bubble, and uniform patterns.

For ultra-thin films we expect $d\ll (L_S, w, L_B, R)$.
In this limit, the energy per sample volume of the stripe configuration has been calculated by Kashuba and Pokrovsky\cite{Kashuba_PRB_1993}. Using our model we obtain an essentially equivalent analytical expression\cite{Saratz_PhD_2009},
\begin{equation}
\label{eq-stripe}
e_{S}=\frac{2J}{L}-\frac{4\lambda d}{L}\left[\ln\left(\frac{2 L}{\pi d}\sin\left(\frac{\pi w}{2 L}\right)\right)+\frac{3}{2}\right] +\frac{h w}{L}\, ,
\end{equation}
except for the constant $\frac{3}{2}$ coming from the slightly different treatment of the dipolar energy and where we have subtracted the energy of the homogeneous, saturated state. For any $h$, the $w$ and $L$ minimizing \eqref{eq-stripe} and the minimized energy can be found analytically. 
For $h=0$, $w$ is equal to $L$ and by minimizing \eqref{eq-stripe} w.~r.~t.\ $L$ we obtain the stripe width in zero field,
\begin{equation}
\label{eq-L0}
L_0 = d\frac{\pi}{2}\exp\left(\frac{2J}{4\lambda d}-\frac{1}{2}\right)\, .
\end{equation}
For the bubbles, the energy per unit volume can be expressed in a series expansion\cite{Saratz_PhD_2009}
\begin{eqnarray}
\label{eq-bubble}
e_{B} = \frac{4 \pi R}{\sqrt{3} L_B^2}\Bigg[2J + R h -4 \lambda d \ln\left(\frac{8 R}{d\sqrt{e}} \right) \nonumber \\ + 6 \pi \lambda d\sum_{k=0}^\infty S_k \left(\frac{R}{L_B}\right)^{3+2k}\Bigg]\, ,
\end{eqnarray}
where the first terms correspond to the energy of a single bubble domain as given by Thiele\cite{Thiele_BellSysTechJ_1969} and the sum accounts for the interaction with the other bubbles in the sample. Note that the coefficients $S_k$ are independent of both, $R$ and $L_B$, and need to be computed only once. Because $\frac{R^2}{L_B^2}<\frac{1}{4}$, the series converges fast and for the precision required here it is sufficient to consider only the first 10 terms of the series. For high magnetic fields  $L_B\gg R$ and in this limit the series vanishes. By minimizing the remaining terms
with respect to $R$ and equating the minimum energy to zero, we can find analytically the critical field above which the uniform state has the lowest energy,
\begin{equation}
\label{eq-hc}
h_C = 32 \lambda \exp\left(-\frac{2J}{4\lambda d} - \frac{3}{2}\right) = \frac{16\pi}{e^{2}}\, \lambda\, \frac{d}{L_0}\, .
\end{equation}
If we substitute the definitions of $h$ and $\lambda$ back into \eqref{eq-hc}, we find a simple scaling expression relating the physical critical field $H_C$ to the characteristic quantities $M_S$ and $L_0$.
\begin{equation}
\label{eq_hml}
H_C = \frac{h_C}{\mu_0 M_S} = \frac{4}{e^{2}}\, M_S\, \frac{d}{L_0}\, .
\end{equation}
In contrast to the stripes, for $0\leq h<h_C$ the energy of the bubbles \eqref{eq-bubble} can only be evaluated numerically.

\begin{figure}
\includegraphics{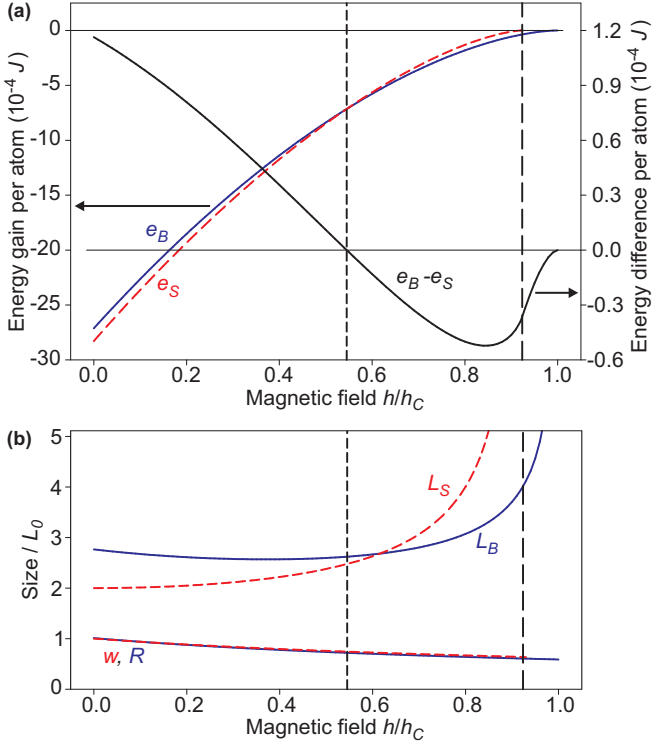}
\caption{(color online). (a) Energies of the optimal stripe and bubble patterns (red-dashed and blue-continues curves, respectively) as a function of the applied field $h$, left scale. Black solid line: energy difference $e_B-e_S$, right scale. The vertical lines indicate the cross-over fields $h_{S\to B}=0.545 h_C$ (short dashed) and $h_{S\to U} = 0.924 h_C$ (long dashed). (b) Size of the optimal domain patterns vs. applied field in units of $L_0$. Upper curves: pattern periods $L_S$ and $L_B$ (red-dashed and blue continuous, respectively). Lower curves: narrow stripe width $w$ and bubble radius $R$ (red-dashed and blue continuous, respectively). Vertical lines as in (a).}
\end{figure}

\begin{figure}
\includegraphics{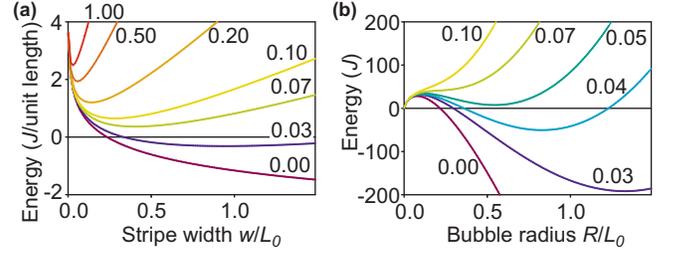}
\caption{(color online). (a) Energy per unit length of a single reversed stripe domain as a function of its width $w$ for different values of the applied field $h$, which is indicated for each curve in units of $\lambda$. (b) Energy of a single reversed bubble domain as a function of its radius $R$ for different values of $h$ in units of $\lambda$, as indicated for each curve.}
\end{figure}

For the following analysis we fix the parameters $J=1$, $d=1$, and $\lambda=0.1$, giving $L_0=141.4$ and $h_C = 0.00481$.
Fig.~8(a) shows the energies of the stripe and bubble patterns (red dashed and blue continuous lines, respectively, left scale) obtained by minimizing the expressions \eqref{eq-stripe} and \eqref{eq-bubble} w.~r.~t.\ ($L$, $w$) and ($L_B$, $R$), respectively.
As it is clear from the graph, both energies increase monotonously for increasing fields and they differ at most by a few percent.
For better clarity, the difference between the two energies ($e_B-e_S$, black solid line, right scale) is also plotted.
The two energies become equal at a critical field $h_{S\to B} = 0.545 h_C$ (vertical short-dashed line), below which the stripes are favored and above which the bubbles have a lower energy.
We find numerically that the value of $h_{S\to B}/ h_C = 0.545$ is independent of the specific choice of $J$, $d$, and $\lambda$, as long as $d/L_0\ll 1$. At a second characteristic field $h_{S\to U}=\frac{e^2}{8} h_C =0.924 h_C$ (vertical long-dashed line), the energy of the stripes becomes equal to the energy of the saturated state, while the bubble pattern still provides an energetic advantage. The value of $h_{S\to U}$ can be computed analytically by taking the limit $L\to\infty$ in \eqref{eq-stripe}, see Eq.~\eqref{eq-eonestripe} below. Hence, all characteristic fields are proportional to $h_c$ and therefore obey the same relation $H\sim M_S d / L_0$, see Eq.~\eqref{eq_hml} and Ref.~\onlinecite{Saratz_PRL_2010}.

In Fig.~8(b) we plot the characteristic lengths of the optimized patterns. The two upper curves indicate the periodicities of the patterns $L_S$ and $L_B$ (red dashed and blue continuous, respectively), while the two lower curves represent the size of the minority domains, $w$ and $R$. Note that the latter are almost equal throughout the entire field range and they remain finite at their respective critical fields $h_{S\to U}$ and $h_C$. The transitions to the saturated state $\frac{w}{L_S}\to 0$ and $\frac{R}{L_B}\to 0$ do not occur through a vanishing numerator, but by a diverging denominator. Nevertheless is the variation in the periodicity rather small throughout most of the field range, except close to the critical fields $h_{S\to U}$ and $h_C$.

\section{Energy barriers and metastability}

Now we want to have a closer look at the mestastability of isolated domains around and above $h_C$ as well as the energy barrier impeding domain merger and splitting in a domain pattern.  

Fig.~9(a) shows the energy (per unit length) of a single reversed stripe domain in an otherwise homogeneous infinite film as a function of its width $w$, for different values of the applied field $h$. The value of $h$ is indicated in units of $\lambda$ for each curve. The expression for the energy per unit length is easily obtained by multiplying Eq.~\eqref{eq-stripe} by the periodicity $L_S=2L$ and the sample thickness $d$ and then taking the limit $L\to \infty$,
\begin{equation}
\label{eq-eonestripe}
E_\mathrm{stripe} = 4 J d- 8\lambda d^2 \left[ \ln\left(\frac{w}{d}\right)+\frac{3}{2}\right]+2 h w d\, .
\end{equation}
For $h>0$ the stripe energy has a shallow minimum at some value $w^*$. For $w\to 0$ Eq.~\eqref{eq-eonestripe} contains an unphysical divergence and is no longer valid. Instead, the stripe reduces to two adjacent domain walls with a total energy corresponding to the associated cost in exchange energy of $4 J d$ per unit stripe length.
This provides a large energy barrier that is independent of $h$ and therefore the isolated stripe domain remains metastable, even for values of $h$ considerably exceeding the equilibrium saturation field $h_C$. Since the stripe domain can not collapse laterally, i.~e.\ by letting $w\to 0$, eventually it will unpin at its ends or split at some point and then contract along its length.

The expression for the energy $E_\mathrm{bubble}$ of an isolated bubble domain\cite{Thiele_BellSysTechJ_1969} with radius $R$ and magnetization $+M_S$ in a homogeneous background of magnetization $-M_S$ is recovered from Eq.~\eqref{eq-bubble} by multiplying the latter by the volume of the unit cell of the bubble lattice, $d\frac{\sqrt{3}}{2}L_B^2$, and then letting $L_B\to\infty$ which corresponds to neglecting the sum,
\begin{equation}
\label{eq-ebub}
E_\mathrm{bubble}=2\pi R d \left[2J -4\lambda d \ln\left(\frac{8 R}{d \sqrt{e}}\right)+R h\right]\, .
\end{equation}
This energy is plotted in Fig.~9(b) for $d=1, \lambda=0.1, J=1$ and some values of the applied field $h$ as indicated in units of $\lambda$ for each curve.
For $h$ small but $>0$, the energy of the bubble domain has two local minima, one at $R=0$ and one at some $R^*$. For $h>h_C$, the state with the bubble domain ($R>0$) becomes metastable, but it is separated form the uniform state ($R=0$) by an energy barrier. However, above a critical field $h_\mathrm{collapse}=\frac{e}{2}h_C$, the local minimum at $R^*$ disappears and the bubble domain inevitably collapses to $R=0$. This behavior is fundamentally different from the stripe domains and we expect that, as long as the domain walls are free to move in the sample, there will be no bubble domains for $h>h_\mathrm{collapse}$, while isolated stripe domains whose ends are pinned at some defect or the sample border may persist up to considerably higher fields.

The reverse process moves the system from the state $R=0$ to $R=R^*$ and corresponds to the nucleation of a single bubble domain. For any $h$ this process is hindered by an intrinsic energy barrier as can be seen in Fig.~9(b). The local maximum of the energy defining the height of the nucleation barrier is found by setting $\partial E_\mathrm{bubble} / \partial R=0$. By using this condition and Eq.~\eqref{eq-hc} in \eqref{eq-ebub} we obtain the following expression for the barrier height $E_B$
\begin{equation}
\label{eq-ebapp}
E_B = 8\pi \lambda d^2 L_0 \frac{R_\mathrm{max}}{L_0}\left(1-\frac{h}{h_C}\frac{4\pi}{e^2}\frac{R_\mathrm{max}}{L_0} \right)\, ,
\end{equation}
where $R_\mathrm{max}$ is the position of the local maximum of the energy, corresponding to the barrier height. The value of the ratio $R_\mathrm{max}/L_0$ is obtained by using the definition of $L_0$, Eq.~\eqref{eq-L0}, in the condition $\partial E_\mathrm{bubble} / \partial R=0$, leading to
\begin{equation}
\label{eq-Rmaxcondition}
\ln\left(\frac{4\pi R_\mathrm{max}}{L_0}\right) - \frac{h}{h_C}\frac{2}{e^2}\frac{4\pi R_\mathrm{max}}{L_0} =0\, .
\end{equation}
Note that the $R_\mathrm{max}/L_0$ is only a function of $h/h_C$. Hence, the energy barrier impeding domain nucleation follows the scaling relation
\begin{equation}
\label{eq-eb}
E_B = \alpha \lambda d^2 L_0
\end{equation}
with the proportionality constant
\begin{equation}
\label{eq-alphadef}
\alpha = 8\pi \frac{R_\mathrm{max}}{L_0}\left(1-\frac{h}{h_C}\frac{4\pi}{e^2}\frac{R_\mathrm{max}}{L_0} \right)
\end{equation}
depending only weakly on $h/h_C$. By solving \eqref{eq-Rmaxcondition} numerically and using the result in \eqref{eq-alphadef} we can compute $\alpha$, e.~g.\ for $h/h_C=+1, 0, -1$ its values are $\alpha = 2.3926, 2.0000, 1.7838$, respectively.
By substituting the definition of $\lambda$ from the appendix A into \eqref{eq-eb} we can relate the barrier height to the experimentally known quantities $M_S$, $d$ and $L_0$,
\begin{equation}
\label{eq-ebappscaling}
E_B = \frac{\alpha}{4\pi} \mu_0 M_S^2 d^2 L_0\, .
\end{equation}
In contrast to the collapse of a domain, nucleating a domain always requires thermal activation to overcome the barrier formed by the exchange interaction which acts like a surface tension, even in absence of pinning at structural defects.

Splitting, or equivalently merging, of domains is hindered by an energy barrier in a similar way, since it implies a transitory increase of the domain wall length that is only partly compensated by a gain in dipolar energy\cite{Cape_JApplPhys_1971}. It seems therefore plausible to assume that the height of the energy barrier impeding domain merger or splitting follows the same scaling behavior \eqref{eq-ebappscaling}, with a different numerical proportionality constant that depends on the exact domain configuration and the details of the process considered. Notice that the same scaling behavior $E_B\sim \lambda L_0$ has been found for the energy of a dislocation in a stripe pattern\cite{Abanov_PRB_1995}.

\end{document}